\documentclass
[showpacs,superscriptaddress,prl,twocolumn,tightenlines,balancelastpage,10pt,a4paper]{revtex4}
\usepackage{amsmath}
\usepackage{amsfonts}
\usepackage{amssymb}
\usepackage{graphicx}
\usepackage{moresize}
\usepackage{epsfig}%

\begin{document}

\title{A millisecond quantum memory for scalable quantum networks}
\author{Bo Zhao}
\thanks{These authors contribute equally to this work}
\affiliation{Physikalisches Institut, Universit\"{a}t Heidelberg,
Philosophenweg 12, D-69120 Heidelberg, Germany}
\author{Yu-Ao Chen}
\thanks{These authors contribute equally to this work}
\affiliation{Physikalisches Institut, Universit\"{a}t Heidelberg,
Philosophenweg 12, D-69120 Heidelberg, Germany} \affiliation{Hefei
National Laboratory for Physical Sciences at Microscale,
Department of Modern Physics, University of Science and Technology
of China, Hefei, 230027, People's Republic of China}
\author{Xiao-Hui Bao}
\affiliation{Physikalisches Institut, Universit\"{a}t Heidelberg,
Philosophenweg 12, D-69120 Heidelberg, Germany} \affiliation{Hefei
National Laboratory for Physical Sciences at Microscale,
Department of Modern Physics, University of Science and Technology
of China, Hefei, 230027, People's Republic of China}
\author{Thorsten Strassel}
\affiliation{Physikalisches Institut, Universit\"{a}t Heidelberg,
Philosophenweg 12, D-69120 Heidelberg, Germany}
\author{Chih-Sung Chuu}
\affiliation{Physikalisches Institut, Universit\"{a}t Heidelberg,
Philosophenweg 12, D-69120 Heidelberg, Germany}
\author{Xian-Min Jin}
\affiliation{Hefei National Laboratory for Physical Sciences at
Microscale, Department of Modern Physics, University of Science
and Technology of China, Hefei, 230027, People's Republic of
China}
\author{J\"{o}rg Schmiedmayer}
\affiliation{Atominstitut der \"Osterreichischen Universit\"aten,
TU-Wien, A-1020 Vienna Austria}
\author{Zhen-Sheng Yuan}
\affiliation{Physikalisches Institut, Universit\"{a}t Heidelberg,
Philosophenweg 12, D-69120 Heidelberg, Germany} \affiliation{Hefei
National Laboratory for Physical Sciences at Microscale,
Department of Modern Physics, University of Science and Technology
of China, Hefei, 230027, People's Republic of China}
\author{Shuai Chen}
\affiliation{Physikalisches Institut, Universit\"{a}t Heidelberg,
Philosophenweg 12, D-69120 Heidelberg, Germany}
\author{Jian-Wei Pan}
\affiliation{Physikalisches Institut, Universit\"{a}t Heidelberg,
Philosophenweg 12, D-69120 Heidelberg, Germany} \affiliation{Hefei
National Laboratory for Physical Sciences at Microscale,
Department of Modern Physics, University of Science and Technology
of China, Hefei, 230027, People's Republic of China}
\date{\today}

\begin{abstract}
Scalable quantum information processing critically depends on the
capability of storage of a quantum state\cite{Briegel98,KLM}. In
particular, a long-lived storable and retrievable quantum memory
for single excitations is of crucial importance to the
atomic-ensemble-based long-distance quantum
communication\cite{DLCZ,Bo07,Zengbing07,jiang07,Collins07}.
Although atomic memories for classical lights\cite{Liu01} and
continuous variables\cite{Julsgaard04} have been demonstrated with
milliseconds storage time, there is no equal advance in the
development of quantum memory for single excitations, where only
around 10 \textbf{$\mu$}s storage time was
achieved\cite{Matsukevich06,Shuai06,SimonJ07,Chou07}. Here we
report our experimental investigations on extending the storage
time of quantum memory for single excitations. We isolate and
identify distinct mechanisms for the decoherence of spin wave (SW)
in atomic ensemble quantum memories. By exploiting the magnetic
field insensitive state, ``clock state", and generating a
long-wavelength SW to suppress the dephasing, we succeed in
extending the storage time of the quantum memory to 1 ms. Our
result represents a substantial progress towards long-distance
quantum communication and enables a realistic avenue for
large-scale quantum information processing.
\end{abstract}


\maketitle

Quantum repeater with atomic ensembles and linear optics has
attracted broad interest in recently years, since it holds the
promise to implement long-distance quantum communication and
distribution of entanglement over quantum networks. Following the
protocol proposed by Duan \textit{et al.}\cite{DLCZ} and the
subsequent improved schemes
\cite{Bo07,Zengbing07,jiang07,Collins07}, significant progresses
have been accomplished, including coherent manipulation of the
stored excitation in one atomic
ensemble\cite{Matsukevich06,Shuai06} and two atomic
ensembles\cite{Felinto06NP,Kuzmich07,Yuan07}, demonstration of
memory-built-in quantum teleportation\cite{Yuao08}, and
realization of a building block of the quantum
repeater\cite{Yuan08,Chou07}. In these experiments, the atomic
ensembles serve as the storable and retrievable quantum memory for
single excitations.

Despite the advances achieved in manipulating atomic ensembles,
long-distance quantum communication with atomic ensembles remains
challenging due to the short coherence time of the quantum memory
for single excitations. For example, to directly establish
entanglement between two memory qubits over a few hundred
kilometers, one needs a memory with a storage time of a few
hundred microseconds. However, the longest storage time  reported
so far is only on the order of 10 $\mu
$s\cite{Matsukevich06,Shuai06,Chou07,SimonJ07}.

It is long believed that the short coherence time is mainly caused
by the residual magnetic field\cite{Felinto06,Choi08}. Thereby,
storing the collective state in the superposition of the
first-order magnetic-field-insensitive
state\cite{Harber02ClockState}, i.e. the ``clock state", is
suggested to inhibit this decoherence mechanism\cite{Felinto06}. A
numerical calculation shows that the lifetime of the collective
excitation stored in the ``clock state" is on the order of
seconds.

Here we report our investigation on prolonging the storage time of
the quantum memory for single excitations. In the experiment, we
find that only using the ``clock state" is not sufficient to
obtain the expected long storage time. We further analyze, isolate
and identify the distinct decoherence mechanisms, and thoroughly
investigate the dephasing of the spin wave (SW) by varying its
wavelength. We find that the dephasing of SW is extremely
sensitive to the angle between the write beam and detection mode,
especially for small angles. Based on this finding, by exploiting
the ``clock state'' and increasing the wavelength of the SW to
suppress the dephasing, we succeed in extending the storage time
from 10 $\mu $s to 1 ms, which for the first time allows for the
quantum memory of single excitations to persist for times
comparable to the propagation of light over 100 kilometers. Our
work has opened the door to the first steps in building a quantum
repeater, and thus provides an essential tool for long-distance
quantum communication.

The architecture of our experiment is depicted in Fig. 1a and 1b. A cold $%
^{87}$Rb atomic ensemble in a magneto-optical trap (MOT) at a temperature of
about 100 $\mu$K serves as the quantum memory. The two ground states $%
|g\rangle$ and $|s\rangle$, together with the excited state
$|e\rangle$ form a $\Lambda$-type system. A bias magnetic field of
about 3.2 G is applied along the axial direction to define the
quantization axis. Note that there are three
pairs of ``clock states" for the ground states of $^{87}$Rb atom, i.e. ($%
|1,1\rangle$, $|2,-1\rangle$), ($|1,0\rangle$,$|2,0\rangle$), and ($%
|1,-1\rangle$, $|2,1\rangle$), where we have defined $|i,j\rangle=|5
S_{1/2}, F=i,m_{F}=j\rangle$. In a timescale of milliseconds, we can use any
of them to store the collective excitation, because the decoherence of the
``clock states" caused by magnetic field is negligible. In our experiment,
we prepare the atoms in $|1,0\rangle$ to exploit the clock state ($%
|g\rangle=|1,0\rangle$, $|s\rangle=|2,0\rangle$). An off-resonant $%
\sigma^{-} $ polarized write pulse with wave vector $\mathbf{k}_{%
\mbox{\ssmall W}}$ is applied to the atomic ensemble along the axial
direction, inducing spontaneous Raman scattering. The Stokes photon with $%
\sigma^{-}$ polarization and wave vector
$\mathbf{k}_{\mbox{\ssmall S}}$ is collected at an angle of
$\theta=3^{\circ}$ relative to the write beam, as in most of the
previous
experiments\cite{Matsukevich06,Kuzmich07,Shuai07,Yuan07,Chou07,Choi08,Felinto06NP}.
The beam waist of the detection mode is about 100 $\mu$m in the
atomic ensemble. Conditional on detecting a Stokes photon, a
collective excited state or a SW is imprinted in the atomic
ensemble\cite{DLCZ}, described by
\begin{equation}
|\psi \rangle =\frac{1}{\sqrt{N}}\sum_{j}e^{i\Delta \mathbf{k}\cdot \mathbf{%
r }_{j}}|g...s_{j}...g\rangle,  \label{psi0}
\end{equation}%
with $\Delta \mathbf{k}=\mathbf{k}_{\mbox{\ssmall
W}}-\mathbf{k}_{\mbox{\ssmall S}}$ the wave vector of the SW, and $\mathbf{r}%
_{j}$ the coordinate of the $j$-th atom. After a controllable delay $\delta
t $, a strong $\sigma^{+}$ polarized read light, counter-propagating with
the write light, converts the collective excitation into an anti-Stokes
photon, which is $\sigma^{+}$ polarized and spatially mode-matched with the
Stokes photon from the opposite direction. The Stokes (anti-Stokes) photon
and the write (read) light are spatially separated.

The quality of the quantum memory can be well characterized by the cross
correlation $g_{\mbox{\ssmall S,AS}}=p_{\mbox{\ssmall
S,AS}}/(p_{\mbox{\ssmall S}}\cdot p_{\mbox{\ssmall AS}})$ (See Methods), with $p_{%
\mbox{\ssmall S}}$ ($p_{\mbox{\ssmall AS}}$) the probability of
detecting a Stokes (anti-Stokes) photon and $p_{\mbox{\ssmall
S,AS}}$ the coincident probability between the Stokes and
anti-Stokes channels. The larger the cross correlation is, the
higher-quality single photon source\cite{Shuai06,Matsukevich06} or
atom-entanglement source\cite{Shuai07,Yuan08} we can acquire. In
the experiment, to evaluate the coherence time of the quantum
memory, we measure the cross correlation as a function of the time
delay, described as\cite{Shuai06}(See Methods)
\begin{equation}
g_{\mbox{\ssmall S,AS}}(\delta t)=1+C\gamma(\delta t),
\end{equation}
with C a fitting parameter, and $\gamma(\delta t)$ the time
dependent retrieval efficiency. Note that $g_{\mbox{\ssmall
S,AS}}>2$ means that the Stokes and anti-Stokes photon are
nonclassically correlated\cite{Chou04,Felinto06}.

The experimental result is given in Fig. 2. Our data shows that
the lifetime is a little bit longer than our previous
results\cite{Shuai06}, where the lifetime is mainly limited by the
residual magnetic field, but far from the theoretical predication
for the ``clock state". From this result, we infer the ``clock
state" can really help to improve the lifetime. But more
importantly, it indicates that, in our experiment, there is
another decoherence mechanism, which dominates when the
decoherence due to magnetic field is suppressed.

We carefully analyze the decoherence mechanism of the quantum
memory and find that the short lifetime could be explained by the
dephasing of the SW induced by atomic random
motion\cite{Mewes05,SimonJ07}. This decoherence mechanism plays an
important role in many previous
experiments\cite{Matsukevich06,Chou07,Shuai07,Yuan07,Kuzmich07,Choi08,Felinto06NP},
but it has not attracted sufficient attention.

The dephasing can be understood as follows. As shown in Fig. 1c, assume a SW
is stored in the atomic ensemble and will be retrieved out after a time
delay $\delta t$. In this interval, each atom randomly moves from one point
to another along the wave vector direction. The internal states or the spin
of the atoms are conserved since collisions can be safely neglected at a low
temperature and density. However, the atomic motion leads to a perturbation
on the phase of the SW. Consequently, the projection of the perturbed SW on
the original one gradually decreases as the delay of the retrieve becomes
longer. In other words, the atomic random motion leads to a random phase
fluctuation in the SW and thus causes decoherence. The timescale of the
dephasing can be estimated by calculating the average time needed for the
atoms to cross $\frac{1}{2\pi}$ of the wavelength of the SW, giving a
lifetime of $\tau_{\mbox{\ssmall D}}\sim\frac{\lambda}{2\pi v_{s}}$, with $%
v_{s}=\sqrt{\frac{k_{\mbox{\ssmall B}}T}{m}}$ the one dimensional average
speed and $\lambda=\frac{2 \pi}{\Delta k}$ the wavelength of the SW. A more
detailed calculation gives $\gamma (\delta t)\sim e^{-\delta t^{2}/\tau_{%
\mbox{\ssmall
D}}^{2}}$, with the lifetime $\tau_{\mbox{\ssmall
D}}=\frac{1}{\Delta k v_{s}}$ (see Methods). In our case, there is an angle $%
\theta$ between $\mathbf{k_{\mbox{\ssmall W}}}$ and $\mathbf{k_{%
\mbox{\ssmall S}}}$, and thus we have $\Delta k=|\mathbf{k}_{\mbox{\ssmall W}%
}-\mathbf{k}_{\mbox{\ssmall S}}|\simeq k_{\mbox{\ssmall
W}}\sin\theta$. For $\theta=3^{\circ}$, a simple calculation gives
$\lambda=15$ $\mu$m and then $\tau_{\mbox{\ssmall D}}=25 $ $\mu$s.
By fitting the data in Fig. 2 with $g_{\mbox{\ssmall S,AS}}(\delta t)=1+C\exp(-\delta t^2/\tau_{%
\mbox{\ssmall D}}^{2})$, we obtain a lifetime of
$\tau_{\mbox{\ssmall D}}=25\pm1 $ $\mu$s, which is consistent with
the theoretical calculation. Note that besides the atomic random
motion, the collisions between atoms may also affect the phase of
the SW. While in our experiment, the effect of collisions is
negligible. The collisions rate can be estimated by $\Gamma\sim
nv_{s}\sigma\simeq 1$ Hz, where the atomic density $n=10^{10}/\mathrm{cm}^3$%
, the s-wave scattering cross section $\sigma=8\pi a^2$ with the
scattering length $a=6$ nm. Thereby, in the time scale of
milliseconds, the collisions can be safely neglected.

To further confirm that the decoherence is mainly caused by the
dephasing induced by atomic motion, we increase the wavelength of
the SW by decreasing the detection angle (see Fig. 1d). In this
way, according to the above model, the dephasing is suppressed and
the storage time will be extended. In our
experiment, we reduce the angle by choosing $\theta =1.5^{\circ }$, $%
0.6^{\circ }$, and $0.2^{\circ }$ and measure the lifetime of the
quantum memory for each configuration. Note that, for $\theta
=0.2^{\circ }$, the two beams with the same polarization can not
be spatially separated, and thereby we use another
\textquotedblleft clock state" ($|g\rangle =|1,1\rangle $,
$|s\rangle =|2,-1\rangle $) by preparing the atoms in $|1,1\rangle
$. In this case, the Stokes (anti-Stokes) photon is $\sigma ^{+}$
($\sigma ^{-}$) polarized. The write (read) and Stokes
(anti-Stokes) lights have orthogonal polarizations and are
separated by a Glan-Laser prism.

The experimental results are displayed in Fig. 3a-3c. As expected,
the dephasing of the SW dominates, when the effect of magnetic
field is inhabited by using the ``clock state". The lifetime
increases from 25 $\mu$s to 283 $\mu$s by reducing $\theta$ or, in
other words, increasing the wavelength of SW. Our results clearly
show that the dephasing of the SW is extremely sensitive to the
small angle between the write beam and Stokes modes, and that the
long-wavelength SW is robust against the
dephasing induced by atomic random motion. Note that, for $%
\theta=0.2^{\circ} $, the data are fitted by taking into account the effect
of loss of atoms. The measured lifetime $\tau_{\mbox{\ssmall D}}$ is shown
in Fig. 3d as a function of angle $\theta$. The solid line is the
theoretical curve $\tau_{\mbox{\ssmall D}}=\frac{1}{\Delta k v_{s}}$, with $%
v_{s}=0.1$ m/s corresponding to a temperature of $T\simeq100$ $\mu$K. The
good agreements between theory and experiment imply that our work provides
an alternative approach to measure the temperature of an atomic ensemble.
Moreover, since the lifetime is only sensitive to the velocity of the atoms
in the interaction region, which is determined by the waist of the detection
mode and is controllable, one can also use our method to measure the
velocity distribution of the atomic ensemble by performing measurement in
different regions.

To further suppress the dephasing and achieve a longer storage time, we use
the collinear configuration ($\theta =0^{\circ }$), where we have the
maximum wavelength of the SW $\lambda \simeq 4.4$ cm and thus $\tau _{%
\mbox{\ssmall D}}\simeq $ 72 ms. In this case, the decoherence due
to loss of atoms, which usually gives a lifetime of a few hundred
microseconds, is isolated as the principal decoherence mechanism.
This mechanism can be estimated by calculating the average time
for the atoms flying out of the pencil shaped interaction region,
where the thermal motion in radial direction dominates. At
temperature $T$, an atomic cloud with a cross section radius
$r_{0}$ expands according to $r^{2}(\delta
t)=r_{0}^{2}+v_{r}^{2}\delta t^{2}$, with the average speed in
radial direction $v_{r}=\sqrt{\frac{2k_{\mbox{\ssmall B}}T}{m}}$.
The retrieval
efficiency can be given by $\gamma (\delta t)=r_{0}^{2}/r^{2}(\delta t)=1/(1+%
\frac{v_{r}^{2}}{r_{0}^{2}}\delta t^{2})$. Thereby, when $\gamma (\tau
_{L})=1/e$, only $1/e$ of the atoms remain in the interaction region, giving
a lifetime of $\tau _{\mbox{\ssmall L}}\simeq \frac{1.31r_{0}}{v_{r}}$. For $%
r_{0}=100$ $\mu $m as the waist of the detection mode and $T=100$
$\mu $K, a direct calculation gives $\tau _{\mbox{\ssmall L}}=950$
$\mu $s, which is much smaller than the dephasing due to atomic
motion and decoherence induced by the magnetic field. The
experimental result is shown in Fig. 4, where the
\textquotedblleft clock state" ($|1,1\rangle $, $|2,-1\rangle $)
is also used. Our data give a lifetime of $\tau _{\mbox{\ssmall
L}}=1.0\pm 0.1$ ms, when the retrieval efficiency has dropped to
$1/e$. The experiment result is in good agreement with the
theoretical estimation.

In our experiment, we have isolated and identified different
decoherence mechanisms of the quantum memory for single
excitations and thoroughly investigated the dephasing of stored SW
by varying its wavelength. Moreover, we have successfully realized
a long-lived quantum memory for single collective excitation by
exploiting the ``clock state" and long-wavelength SW. The storage
time of 1 ms is 30 times longer than the best result reported so
far\cite{Matsukevich06}, and is long enough for photons
transmission over 100 kilometers. In our experiment, the coherence
time of the quantum memory is limited by the decoherence due to
loss of atoms, which can be suppressed by lowering the temperature
via optical molasses. A storage time of 3 ms is achievable by
reducing the temperature to 10 $\mu$K. This will be the upper
limit for the atomic memory in MOT, since longer storage time is
prohibited by the free falling of the atoms under gravity. Further
improvement might be achieved by trapping the atoms in an optical
dipole trap\cite{Felinto06,Grimm00}, where the decoherence due to
loss of atoms and the dephasing induced by atomic random motion
can both be suppressed. In this case, the principal decoherence
mechanism is the diffusion caused by collisions, which will give a
lifetime of a few tens of milliseconds. To inhibit the
collision-induced diffusion, one has to trap
the atoms in a deep optical lattice\cite{Greiner02} or use solid state system%
\cite{Longdell05}, where each atom is tightly confined in a single site and
collisions are avoided. The optical lattice has the potential to store the
collective excitation for a few tens of seconds, which will reach the
requirement in the storage time for a robust and efficient quantum repeater
with atomic ensembles\cite{Sangouard08}. The idea presented in this work can
also be applied to the quantum memory based on electromagnetically induced
transparency\cite{Fleischhauer00,Kuzmich05,Choi08}. By using the same method
as in our experiment, a storage time of a few hundred microseconds can be
expected.

Our work opens up the exciting possibility to implement many tasks
of quantum information processing. Combined with the techniques
developed in recent years, one can implement a high-quality
on-demand single-photon source, deterministic preparation of
multi-qubit entanglement, generation of entanglement between two
remote atomic memory qubits over a few hundred kilometers, and
even construction of long-lived quantum nodes for quantum
repeater. More generally, our work presents an experimental
investigation on the decoherence of the SW at single quanta level.
It is clearly shown that a long-wavelength SW is robust against
dephasing. Besides, our work also provides an approach to measure
the temperature or the velocity distribution of an atomic
ensemble. Furthermore, since the decoherence of the SW is
controllable, one can measure certain important physical
quantities by introducing additional physical mechanisms. For
example, when performing experiments in optical dipole trap, the
lifetime is determined by collision between atoms\cite{Mewes05}.
Thereby, the s-wave scattering cross section or scattering length
might be measured using our approach.

We acknowledge M. Fleischhauer and Y. J. Deng for useful
discussions. This work was supported by the Deutsche
Forschungsgemeinschaft (DFG), the Alexander von Humboldt
Foundation, the European Commission through the Marie Curie
Excellence Grant, the ERC Grant, the National Fundamental Research
Program (Grant No.2006CB921900), the CAS, and the NNSFC.\\
$*$These authors contributed equally to this work.

\textbf{Correspondence} and requests for materials should be
addressed to Y.A.C~(email:yuao@physi.uni-heidelberg.de) or
J.W.P~(email: jian-wei.pan@physi.uni-heidelberg.de).
\bibliographystyle{plain}
\bibliography{memory}

\clearpage

\textbf{Fig. 1.} (a) Schematic view of the experiment. The atoms
are initially prepared in $|g\rangle$. A weak $\sigma^{-}$
polarized write pulse is applied to generate the SW and Stokes
photon via spontaneous Raman transition
$|g\rangle\rightarrow|e\rangle\rightarrow|s\rangle$. The Stokes
photon are detected at an angle of $\theta$ relative to the write
beam. After a controllable delay, a strong $\sigma^{+}$ polarized
read light induces the transition
$|s\rangle\rightarrow|e\rangle\rightarrow|g\rangle$, converting
the SW into an anti-Stokes photon. (b) The structure of atomic
transitions ($^{87}$Rb) under a weak magnetic field. The left
panel corresponds to the experiment with ($|1,0\rangle,
|2,0\rangle$). The right one corresponds to the experiment with
($|1,1\rangle, |2,-1\rangle$). The photons emitted in undesired
transitions are filtered by polarization filter and filter cell.
(c) Illustration of the SW dephasing induced by atomic random
motion. The blue curve represents the SW initially stored in the
quantum memory. The atoms randomly move along the wave vector
direction, resulting in a phase fluctuation. The perturbed SW is
represented by the red curve. (d) The wavelength of the SW can be
controlled by changing the detection configuration. In the
collinear case, we have the maximum wavelength.

\bigskip

\textbf{Fig. 2.} The cross correlation $g_{\mbox{\ssmall S,AS}}$ versus the
storage time $\delta t$ for ($|1,0\rangle, |2,0\rangle$) at $%
\theta=3^{\circ} $. The data are fitted by using $g_{\mbox{\ssmall S,AS}%
}(\delta t)=1+C\exp(-\delta t^2/\tau_{\mbox{\ssmall D}}^{2})$. Our data give
a lifetime of $\tau_{\mbox{\ssmall D}}=25\pm1$ $\mu$s, which is much less
than the theoretical estimation for the ``clock state". Error bars represent
statistical errors.

\bigskip

\textbf{Fig. 3.} The cross correlation $g_{\mbox{\ssmall S,AS}}$ versus the
storage time $\delta t$ for different angles (a)-(c) and the measured
lifetime $\tau_{\mbox{\ssmall D}}$ as a function of detection angle $\theta$
(d). Panels (a) and (b) are for ($|1,0\rangle, |2,0\rangle$) at $%
\theta=1.5^{\circ}$ and $0.6^{\circ}$, respectively. The data are fitted by
using $g_{\mbox{\ssmall S,AS}}(\delta t)=1+C\exp(-\delta t^2/\tau_{%
\mbox{\ssmall D}}^{2})$, where $\tau_{\mbox{\ssmall D}}$ is the lifetime due
to dephasing. Panel (c) is for ($|1,1\rangle, |2,-1\rangle$) at $%
\theta=0.2^{\circ}$. In this case we take into account the effect of loss of
atoms and fit the data by using $g_{\mbox{\ssmall S,AS}}(\delta
t)=1+C\exp(-\delta t^2/\tau_{\mbox{\ssmall D}}^{2})/(1+A\delta t^2)$, with $%
A $ the fitting parameter obtained from the collinear configuration. The
fitted lifetime for each case is: (a) $\tau_{\mbox{\ssmall
D}}=61\pm2$ $\mu$s, (b) $\tau_{\mbox{\ssmall D}}=144\pm9$ $\mu$s, (c) $\tau_{%
\mbox{\ssmall D}}=283\pm18$ $\mu$s. The first data are a little bit higher
than the fitted curves, which might be caused by the imperfection in the
pumping process. By reducing the angle, the lifetime is increased from 25 $%
\mu$s to 283 $\mu$s, which implies the decoherence is mainly caused by the
dephasing induced by atomic random motion. Panels (d) depicts the measured
lifetime $\tau_{\mbox{\ssmall D}}$ as a function of detection angle $\theta$%
, where the horizontal error bars indicate measurement errors in the angles.
The solid line is the theoretical curve with $T\simeq100$ $\mu$K. The
experimental results are in good agreement with the theoretical
predications. The vertical error bars indicate statistical errors.

\bigskip

\textbf{Fig. 4.} The cross correlation $g_{\mbox{\ssmall S,AS}}$ versus the
storage time $\delta t$ for $\theta=0^{\circ}$ and ($|1,1\rangle,
|2,-1\rangle$). The data are fitted by using $g_{\mbox{\ssmall S,AS}}(\delta
t)=1+\frac{C}{1+A \delta t^{2}}$, with $A$ the fitting parameter. Our data
give a lifetime of $\tau_{\mbox{\ssmall L}}=1.0\pm0.1$ ms, when the
retrieval efficiency $\gamma(\delta t)=\frac{1}{1+A \delta t^2}$ has dropped
to $1/e$. Error bars represent statistical errors.

\clearpage

\clearpage
\begin{figure}[tbp]
\begin{center}
\epsfig{file=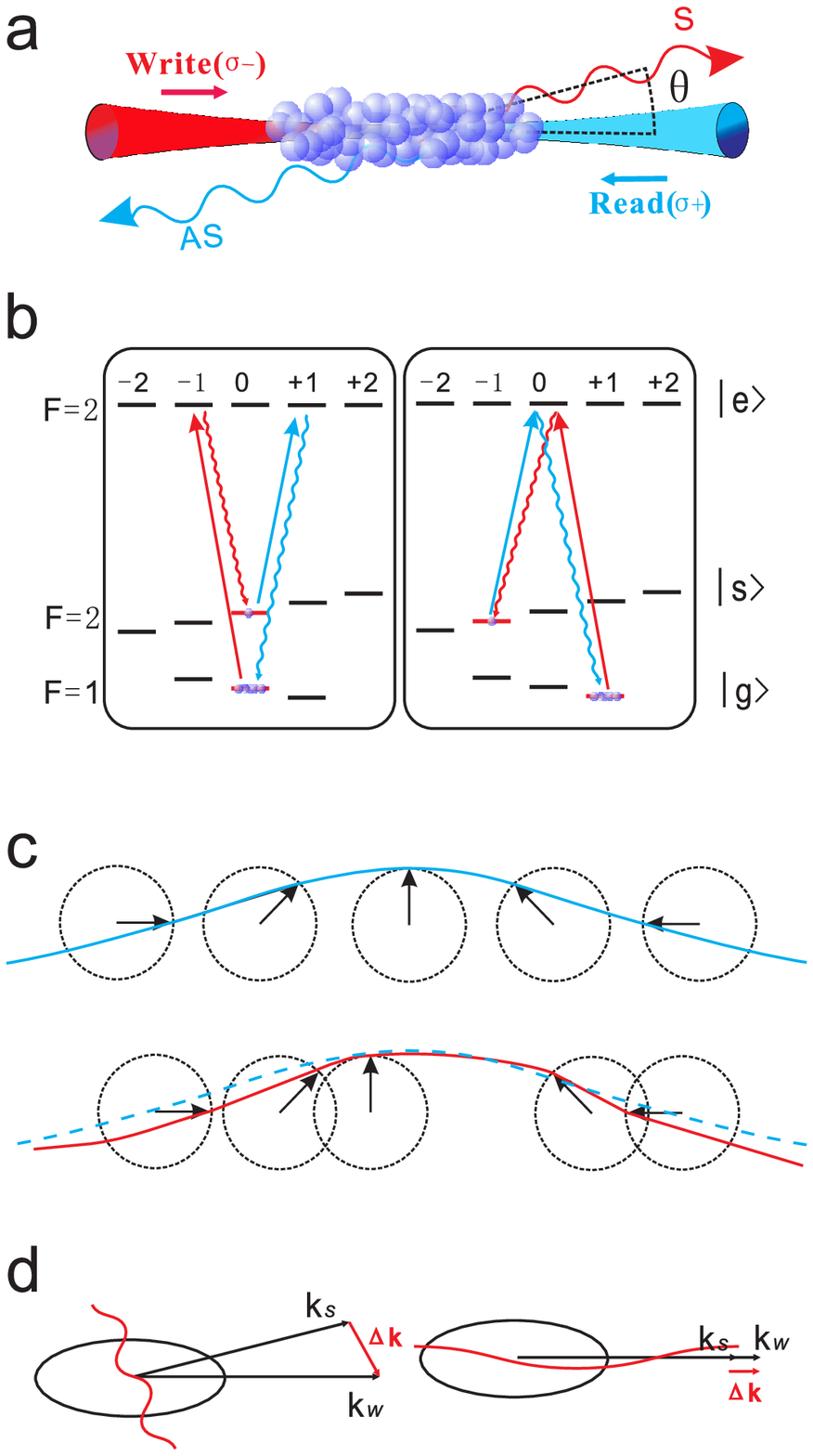,width=10cm}
\end{center}
\caption{}
\end{figure}

\clearpage
\begin{figure}[tbp]
\begin{center}
\epsfig{file=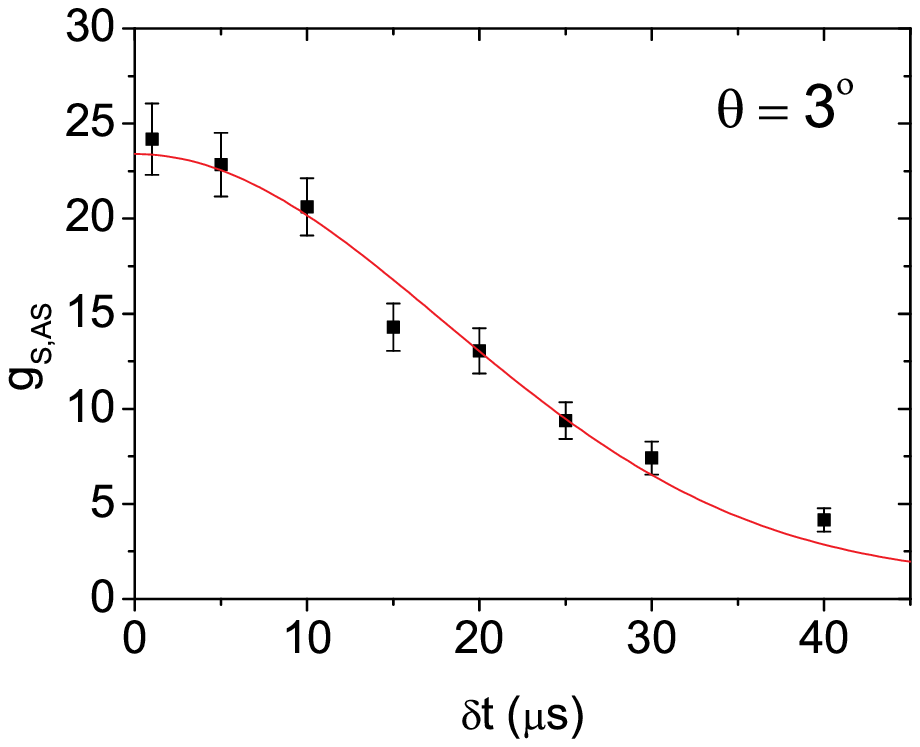,width=14cm}
\end{center}
\caption{}
\end{figure}

\clearpage
\begin{figure}[tbp]
\begin{center}
\epsfig{file=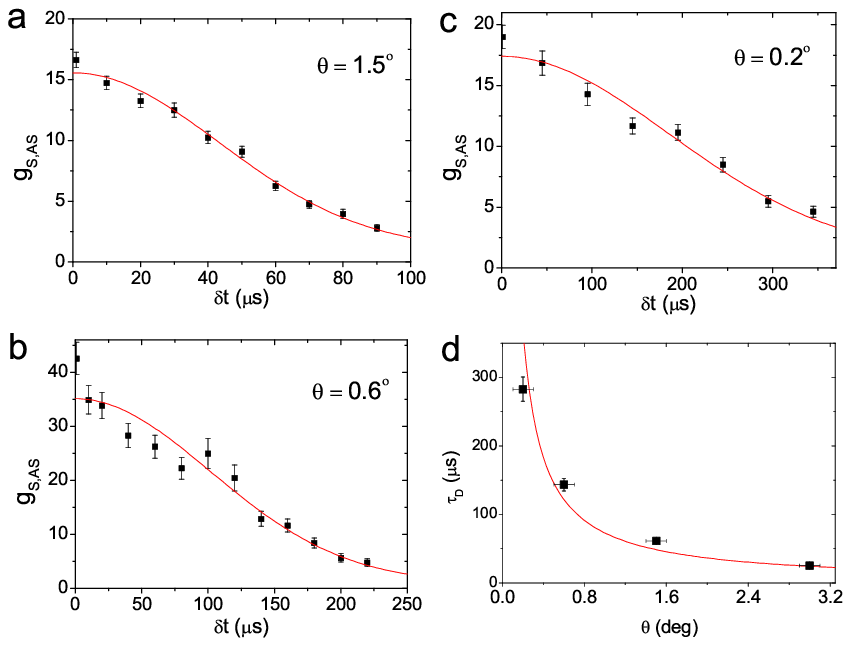,width=16cm}
\end{center}
\caption{}
\end{figure}

\clearpage
\begin{figure}[tbp]
\begin{center}
\epsfig{file=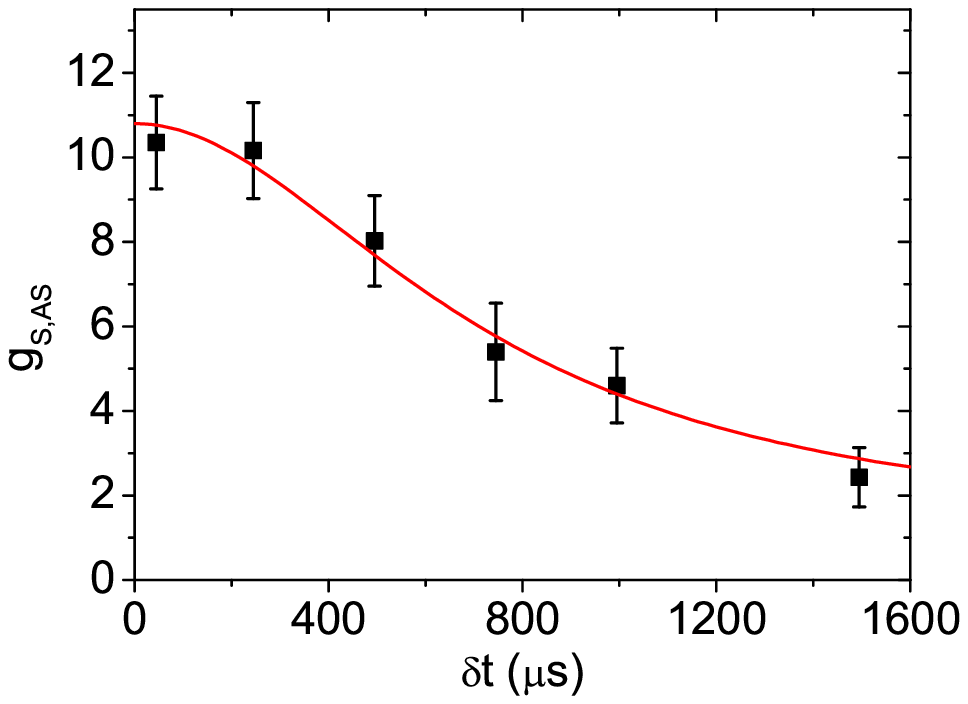,width=16cm}
\end{center}
\caption{}
\end{figure}

\clearpage

\section*{Methods}

\noindent \textbf{Dephasing of Spin Wave Induced by Atomic Random Motion.}
Assume the $j$-th atom moves to $\mathbf{r}_{j}(\delta t)=\mathbf{r}_{j}+%
\mathbf{v}_{j}\delta t$ after a storage time of $\delta t$. The collective
state or spin wave (SW) freely evolves to
\begin{equation}
|\psi_{\mbox{\ssmall D}}\rangle=\frac{1}{\sqrt{N}}\sum_{j} e^{i \Delta
\mathbf{k}\cdot \mathbf{r}_{j}(\delta t)}|g...s_{j}...g\rangle,
\end{equation}
where we have neglected the effect of magnetic field for simplicity. The
retrieval efficiency is given by the overlap between the original SW and the
perturbed one,
\begin{equation}
\gamma(\delta t)\sim|\langle\psi|\psi_{\mbox{\ssmall
D}}\rangle|^{2}=|\frac{1}{N}\sum_{j}e^{i\Delta \mathbf{k}\cdot\mathbf{v}%
_{j}\delta t}|^{2}=|\int f(\mathbf{v})e^{i\mathbf{\Delta k}\cdot\mathbf{v}%
\delta t}d\mathbf{v}|^{2}
\end{equation}
with $f(\mathbf{v})$ the velocity distribution. Assume $f(\mathbf{v})\sim
e^{-\frac{m \mathbf{v}^{2}}{2 k_{\mbox \ssmall
B}T}}$ is a Boltzmann distribution at temperature $T$. Integrating over all
possible velocity, we obtain $\gamma (\delta t)\sim e^{-\delta t^{2}/\tau_{%
\mbox{\ssmall D}}^{2}}$, with the lifetime $\tau_{\mbox{\ssmall D}}=\frac{1}{%
\Delta k v_{s}}$.

\noindent \textbf{Cross Correlation Function.} The quality of the
quantum memory can be well characterized by the cross correlation $g_{%
\mbox{\ssmall S,AS}}=p_{\mbox{\ssmall S,AS}}/(p_{\mbox{\ssmall
S}}\cdot p_{\mbox{\ssmall AS}})$. For $g_{\mbox{\ssmall S,AS}}\gg
1,$when the atomic ensemble is used as a single photon
source\cite{Shuai06,Matsukevich06}, the autocorrelation
of the heralded single photons can be estimated by $\alpha \simeq 4/(g_{%
\mbox{\ssmall S,AS}}-1)$. When the atomic ensemble is used to
prepare atom-photon entanglement source\cite{Shuai07,Yuan08}, the
violation of the CHSH-type Bell
inequality can be approximated by $S\simeq 2\sqrt{2}(g_{\mbox{\ssmall S,AS}%
}-1)/(g_{\mbox{\ssmall S,AS}}+1)$. Thereby, a large cross
correlation function indicates a high-quality quantum memory. The
nonclassical correlation between Stokes and anti-Stokes fields is
characterized by the
violation of Cauchy-Schwarz inequality $g_{\mbox{\ssmall S,AS}}^{2}\leq g_{%
\mbox{\ssmall S,S}}g_{\mbox{\ssmall AS,AS}}$, with
$g_{\mbox{\ssmall S,S}}$ and $g_{\mbox{\ssmall AS,AS}}$ are the
autocorrelation of Stokes and anti-Stokes field. In ideal case,
$g_{ \mbox{\ssmall S,S}}=g_{\mbox{\ssmall AS,AS}}=2$, and in
practice, they are usually smaller than 2 due to the back ground
noise\cite{Chou04}. Thereby, one can infer when $g_{\mbox{\ssmall
S,AS}}>2$, the Stokes and anti-Stokes field are nonclassically
correlated\cite{Felinto06}. In the experiment, we measure the
decay of the cross correlation to evaluate the lifetime of the
quantum memory. By neglecting the noise in Stokes channel, we
have\cite{Shuai06}
\begin{eqnarray}
p_{\mbox{\ssmall
S}} &=&\chi \eta _{\mbox{\ssmall
S}}, \\
p_{\mbox{\ssmall AS}} &=&\chi \gamma (\delta t)\eta
_{\mbox{\ssmall AS}}+B\eta _{\mbox{\ssmall
AS}}, \\
p_{\mbox{\ssmall S,AS}} &=&\chi \gamma (\delta t)\eta _{\mbox{\ssmall S}%
}\eta _{\mbox{\ssmall AS}}+p_{\mbox{\ssmall S}}p_{\mbox{\ssmall
AS}},
\end{eqnarray}
with $\chi $ the excitation probability, $\gamma (\delta t)$ the
time dependent retrieval efficiency, $\eta _{\mbox{\ssmall
S}}(\eta _{\mbox{\ssmall AS}})$ the overall detection efficiencies
in the Stokes (anti-Stokes) channel, and $B$ the background noise
in anti-Stokes channel. Thereby the decay of the cross correlation
function can be approximated by,
\begin{equation}
g_{\mbox{\ssmall S,AS}}(\delta t)=1+\frac{\gamma (\delta t)}{\chi
\gamma (\delta t)+B}\simeq 1+C\gamma (\delta t),
\end{equation}%
where $C$ is a fitting parameter. In our experiment,
$p_{\mbox{\ssmall S}}\simeq0.003$ and the cross correlation is
comparable with previous works\cite{Shuai06}. For small angles and
in the collinear regime, the cross correlation is lower because of
the relatively large background noise due to the write and read
beams. Note that by reduing the power and waist of write and read
beams, and improving the filtering techniques, a high cross
correlation of about 100 can be achieved in the collinear
configuration\cite{Shuai06}.

\bigskip

\noindent \textbf{Experimental Details.} In the experiment, the MOT is
loaded for 20 ms at a repetition rate of 40 Hz. The trapping magnetic field
and repumping beams are then quickly switched off. After 0.5 ms, the bias
magnetic field is switched on, whereas the cooler beams stay on for another
0.5 ms before being switched off to prepare the atoms in the $%
|5S_{1/2},F=1\rangle $ ground state. Then, within another 4 ms,
experimental trials (each consisting of pumping, write and read
pulses ) are repeated with a controllable period depending on the
desired retrieval time. In order to optically pump the atoms to
the desired sub-level, we switch on two pumping beams in each
experimental trial before write and read process: one couples the
transition $|5S_{1/2},F=2\rangle \rightarrow |5P_{3/2},F^{\prime
}=2\rangle $ with linear polarization for 2 $\mu s$, and the other
couples the transition $|5S_{1/2},F=1\rangle \rightarrow
|5P_{1/2},F^{\prime }=1\rangle $ for 1.7 $\mu s$, which is
linearly ($\sigma ^{+}$) polarized for $|1,0\rangle $
($|1,1\rangle $). From the experimental result, we estimate more
than $80\%$ of the atoms are prepared to the desired state.

In our experiment, more than $10^8$ $^{87}$Rb atoms are collected by the MOT
with an optical depth of about 5 and a temperature of about 100 $\mu$K. The
write pulse with a detuning of $\Delta =20$ MHz and a beam diameter of about
400 $\mu $m is applied to generate the SW and the Stokes photon. The Stokes
mode is coupled into a single-mode fiber (SMF) and guided to a single-photon
detector. After a controllable delay, the strong read pulse with a detuning
of $\Delta =6$ MHz is applied to retrieve the SW. The stored SW is converted
into the anti-Stokes photon, which is also coupled into a SMF and detected
by a single-photon detector.

In the experiment, for $\theta =3^{\circ },1.5^{\circ }$ and $0.6^{\circ }$,
the Stokes (ant-Stokes) photon and the write (read) light can be spatially
separated and thus we can choose any of the three pairs of \textquotedblleft
clock states". Because the retrieval efficiency is proportional to the
coupling strength of the transition $|e\rangle \rightarrow |g\rangle $, we
choose the clock state ($|1,0\rangle $, $|2,0\rangle $) to get higher
retrieval efficiency. While for smaller angles $\theta =0.2^{\circ }$ and $%
0^{\circ }$, the two beams with the same polarization cannot be spatially
separated, and thereby we have to use the other two \textquotedblleft clock
states". In this case, we choose the \textquotedblleft clock state" ($%
|1,1\rangle $, $|2,-1\rangle $), since the energy level $|1,1\rangle $ is
lower than $|1,-1\rangle $ under the bias magnetic field and the pumping
effect is better. The overall retrieval efficiency, including transmission
efficiency of filters and optical components, the coupling efficiency of the
fiber coupler, and the detector quantum efficiency, are about $2\%$ for ($%
|1,0\rangle $, $|2,0\rangle $) at $\theta =3^{\circ },1.5^{\circ }$ and $%
0.6^{\circ }$, and $1\%$ for ($|1,1\rangle $, $|2,-1\rangle $) at
$\theta =0.2^{\circ }$, and $0.8\%$ for ($|1,1\rangle $,
$|2,-1\rangle $) at $\theta =0^{\circ }$. The retrieval efficiency
at $\theta =0^{\circ }$ is a little bit lower than at
$\theta=0.2^{\circ }$, because one more etalon was used to filter
the excitation beams. These correspond to $15\%$ and $10\%$ of
original
retrieval efficiency for ($|1,0\rangle $, $|2,0\rangle $) and ($|1,1\rangle $%
, $|2,-1\rangle $), respectively. The low overall retrieval
efficiency is caused by the transmission loss, the mode mismatch,
the imperfect pumping, and the imperfect polarization of the write
and read light.

\end{document}